**Enhanced readout contrast of V2 ensembles in 4H-SiC through resonant optical excitation**

Infiter Tathfif [1,2], and Samuel G. Carter [1,*]

[1] *Laboratory for Physical Sciences, College Park, MD 20740, USA*

[2] *Department of Electrical and Computer Engineering, University of Maryland, College Park, MD 20742, USA*

**ABSTRACT.** The V2 silicon vacancy defect in 4H-SiC has emerged as a promising system for quantum technologies due to its favorable optical and spin properties and the advantages of the SiC host. However, the readout contrast – an important benchmark for quantum sensing – of V2 ensembles for optically-detected magnetic resonance (ODMR) is relatively low, usually <1% at room temperature. To overcome this challenge, we resonantly excite the V2 ensembles at cryogenic temperatures and compare the results with the off-resonant case. We report a maximum ODMR contrast of 50% with only 2 µW of resonant laser power, almost 100 times improvement over off-resonant excitation. We attribute this high readout contrast to a subset of V2 centers that have one spin-selective optical transition resonant with the laser. The ODMR contrast decreases with temperature, approaching the non-resonant contrast by 60 K, likely due to broadening of the optical transition linewidths. We achieve a maximum sensitivity of 100 nT/√Hz with a resonant laser power of 300 µW, while 100 times more non-resonant excitation power is needed to achieve comparable sensitivity.

## I. INTRODUCTION

Color centers in wide-bandgap materials have garnered significant interest over the past two decades for their optical and spin-state properties, such as stable single photon emission, highly coherent spin states at room temperature, and high-fidelity optical initialization and readout [1–10]. Leveraging these features, with the added benefit of nanoscale control, researchers have utilized these color centers for a myriad of applications in quantum technologies, for example, in communication, computing, and sensing [11–24]. Nitrogen-vacancy (NV) centers in diamond, composed of a substitutional nitrogen atom next to a vacancy in the carbon lattice, are currently the prime candidate for such spin-defects due to their well-understood physics and extensive research [21–30]. However, difficulty of diamond fabrication, expense of high-quality diamond samples, and small sample sizes limit the efficacy and versatility of this system [31]. Therefore, there is a compelling need for alternative spin-defect platforms.

* Corresponding author: sgcarte@lps.umd.edu

Recently, color centers in SiC have gained considerable traction within the scientific community for several reasons. Firstly, the point defects in SiC exhibit long spin coherence times comparable to NV centers [32–35]. Secondly, SiC is compatible with industry standard CMOS technology, has mature fabrication techniques, and high-quality epitaxial films are readily and commercially available at a lower price than diamonds [36,37]. Finally, the ability to control doping in SiC wafers, a challenge for diamond films, offers experimental freedom [31]. The most widely studied color centers in SiC include negatively charged silicon vacancy ($V_{Si}^-$), neutral divacancy ($V_C V_{Si}$) and negatively charged NV centers ($N_C V_{Si}^-$) [36]. Due to the ease of defect production, considerable research has been done on the $V_{Si}^-$ defect. In 4H-SiC, the $V_{Si}$ can be located at either a hexagonal (V1) or cubic (V2) site [38,39]. V2 centers remain the leading choice for research as they exhibit optically-detected magnetic resonance (ODMR) at room temperature.

Despite having promising optical and spin properties, the V2 ensembles in 4H-SiC have relatively weak ODMR contrast, defined as the % change in photoluminescence (PL) between when spin transitions are driven or not driven. This contrast, which is inversely proportional to magnetic sensitivity, is typically about ~0.5% for continuous wave (cw) ODMR with off-resonant excitation [33]. In contrast, the NV ensembles in diamond outperform the V2 ensembles by having a ~2% ODMR contrast [40]. To address this issue, one approach is to use resonant excitation of V2 ensembles at the zero-phonon line (ZPL) at cryogenic temperatures. Previous work with resonant excitation demonstrated high ODMR contrast for single defects [35] as well as V1 ensembles in 4H-SiC [41] and $V_{Si}$ ensembles in 6H-SiC [42]. However, these studies neither characterized ODMR contrast based on laser powers, PL, and temperature nor investigated V2 ensembles in 4H-SiC.

In this paper, we explore resonant excitation of V2 ensembles in 4H-SiC and achieve a maximum ODMR contrast of 50%. Furthermore, we make in-depth comparisons of our findings with off-resonant excitation. Section II details the 4H-SiC sample and experimental methods. Section III is divided into the following sections: Section IIIA describes the PL spectra for the $V_{Si}$ ensembles with a photoluminescence excitation (PLE) scan of the V2 center; Section IIIB presents results on ODMR of V2 for non-resonant and resonant laser excitation; Section IIIC describes the effects of laser linewidth modulation and temperature variation on ODMR contrast; and Section IIID calculates the magnetic sensitivity for both excitation types. In Section IV, we explain the underlying mechanism of resonant excitation of V2 ensembles. Section V summarizes the results.

## II. EXPERIMENTAL METHODS

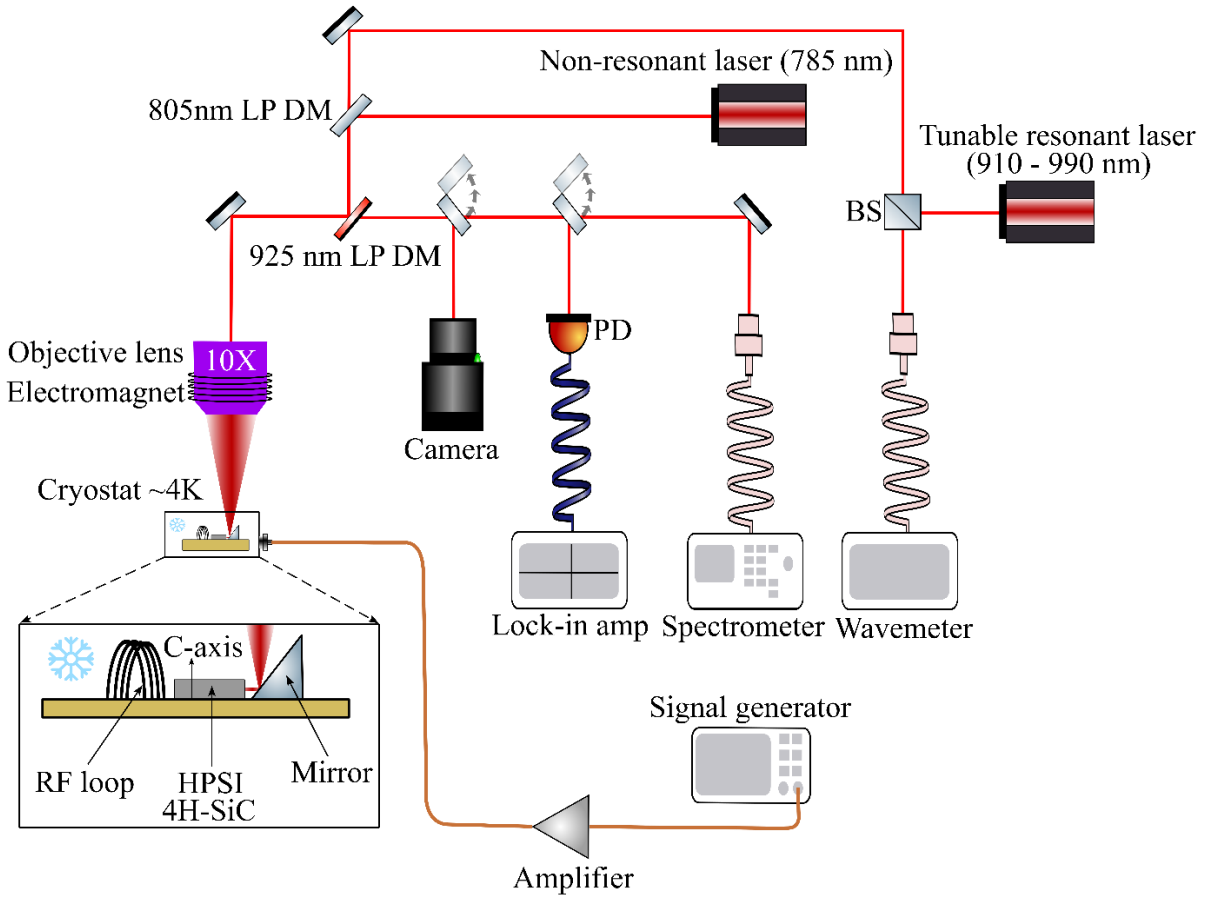

Figure 1: Overview of the experimental setup for resonant excitation of the V2 ensembles in 4H-SiC. The microscope objective focuses the laser beams into the prism, excites the edge of the sample and collects PL. The collected PL can be sent to a camera, photodiode, or spectrometer, with BNC cables and multimode optical fibers shown in blue and pink, respectively. The signal generator with an amplifier supplies the RF drive to a small coil for cw ODMR. The cryostation, signal generator and amplifier are connected through SMA cables (brown).

The detailed experimental setup is illustrated in Fig. 1. We use a commercially available high-purity semi-insulating (HPSI) 4H-SiC wafer from Silicon Valley Microelectronics (SVM). The SVM wafer is ~500 μm thick and has been irradiated with 1 MeV electrons at a dose of $1 \times 10^{19}$ cm$^{-2}$ with no annealing. The wafer is then diced in-house in 3 mm × 3 mm squares, and we place one of the samples into an enclosed Montana Instruments Cryostation. Based on Ref. [43], we estimate the density of $V_{Si}$ to be $10^{16} - 10^{17}$ cm$^{-3}$.

We employ two cw lasers in the setup for cryogenic (~4 – 5 K) microscopy: a tunable laser (TOPTICA CTL 950) excites the V2 centers resonantly at ~916 nm, and a 785 nm laser (Cobolt 06-MLD) excites V1 and V2 centers non-resonantly. A small portion of the tunable laser is branched off with a beam splitter (BS) to send to the high-resolution wavemeter WS6 from HighFinesse. Both laser beams are sent to a Thorlabs 10X microscope objective (0.5 NA) through their respective long-pass (LP) dichroic mirrors (DMs). The objective lens focuses the beams into a 3 mm silver-coated, N-BK7 90º mirror from Edmund Optics and excites the edge of the sample with a laser spot size of approximately 10 μm. The laser polarization is aligned to the c-axis of the SiC. This geometry ensures maximum zero phonon line (ZPL) absorption and emission since the dipole of the V2 transition is along the c-axis [44].

A handmade coil of 5 loops and ~3 mm diameter supplies the RF magnetic driving field. The RF loop is secured in a slot within the sample mount and placed close to the diced square itself. The ends of the RF loop are connected to a coax cable that is fed through the cryostation to the Agilent 81150 signal generator and Mini-Circuits ZHL-10W amplifier, which are used to supply and amplify the RF signal, respectively. The RF signal is modulated with a 900 Hz square wave through the signal generator directly for cw ODMR measurements. We also use an in-house electromagnet made of approximately 40 turns, surrounding the microscope objective. We estimate a magnetic field of 0.5 mT along the c-axis by supplying 1.5 A of constant current using a Keithley 2460 source meter.

The PL is collected through the same objective lens and can be directed to different devices using flip mirrors. We use a short-wave infrared (SWIR) camera from Sensors Unlimited (640CSX) for imaging the PL. We can also focus the PL into a Thorlabs APD410C InGaAs avalanche photodetector (APD). With an LP filter in front of the APD, the range of detection is limited to 950 nm – 1600 nm. The APD signal is then sent to a Stanford Research Systems 500 kHz lock-in amplifier (SR860). Moreover, we can also analyze the PL spectrum through a Princeton Instruments spectrometer HRS-750 and a Princeton Instruments Blaze camera (400 nm – 1050 nm).

## III. RESULTS

### A. Photoluminescence (PL) spectrum and photoluminescence excitation (PLE) scan

The PL spectrum of the SVM HPSI 4H-SiC at ~4 K is displayed in Fig. 2(a). The ZPLs of V1 and V2 with broad phonon sidebands (PSBs) are identified. Since this study is focused on the V2 center, we perform a PLE scan around the ZPL of V2. The tunable laser is scanned around the ZPL of V2 center while measuring the integrated PL with the APD. The wavelength of the tunable laser is calibrated by the high-resolution WS6 wavemeter. Fig. 2(b) displays the PLE scan of the V2 centers. The peak of the PLE scan is at ~916.49 nm with a full-width half maximum (FWHM) of ~0.13 nm. Resonant ODMR is performed with the tunable laser at this peak.

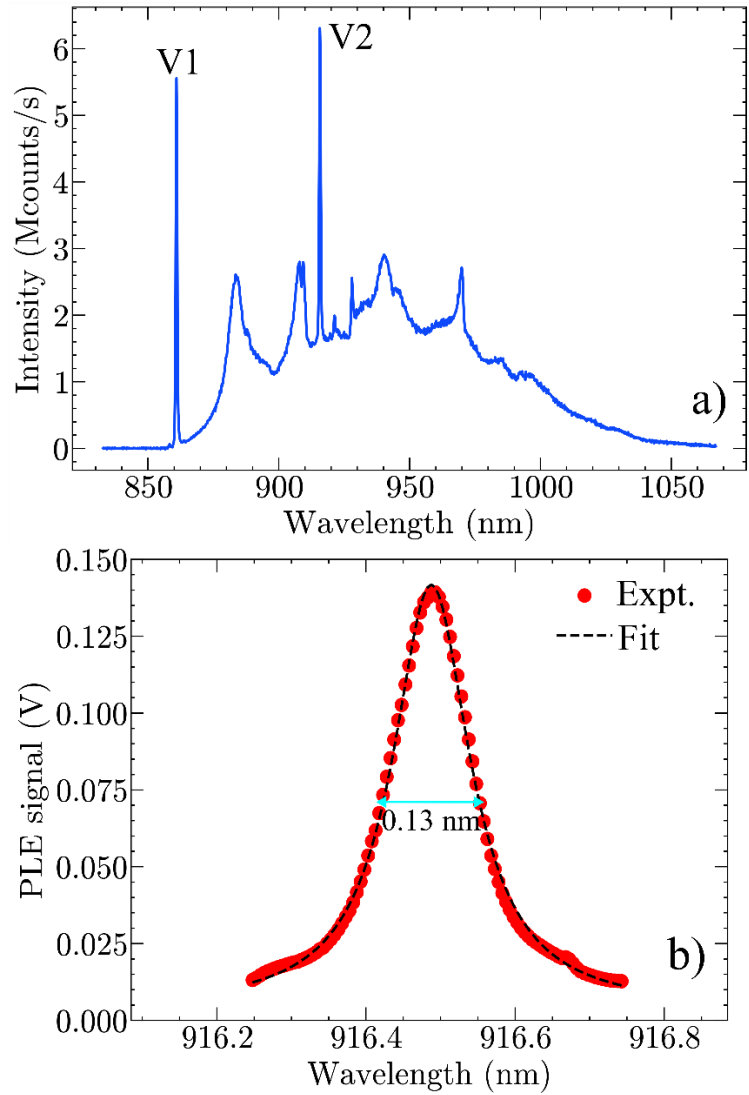

Figure 2: (a) PL spectrum and (b) PLE scan of the HPSI 4H-SiC. PL spectrum exhibits ZPLs of V1 and V2 defects with broad PSBs. The peak of the PLE scan identifies the resonance of the V2 defects to perform resonant ODMR.

## B. Optically-detected magnetic resonance (ODMR) spectra and contrast analysis

The V2 ground state is a $S = 3/2$ system and has the following Hamiltonian:

$$H = g\mu_B \boldsymbol{B} \cdot \boldsymbol{S} + DS_z^2 + H_{hf}, \qquad (1)$$

where, $g$ is the electronic $g$ factor, $\mu_B$ is the Bohr Magneton, $\boldsymbol{B}$ is the magnetic field vector, $\boldsymbol{S} = (S_x, S_y, S_z)$ is the spin-3/2 operator, $2D = 70$ MHz is the zero-field splitting, $z$ is along the c-axis, and $H_{hf}$ is the hyperfine Hamiltonian. Fig. 3(a) plots the energy level diagram of the $S = 3/2$ system of the V2 ground state. The red arrows indicate magnetic dipole allowed $\Delta m_s = \pm 1$ transitions.

Fig. 3(b) plots the cw ODMR spectra for two different RF powers (3 dBm and 16 dBm) with off-resonant laser excitation. To calculate the contrast, we divide the ODMR signal by the PL obtained with no RF drive. As previously observed, the most intense ODMR signals occur for -3/2 → -1/2 and +1/2 → +3/2 transitions [32,45]. Therefore, we label the two most prominent peaks in Fig. 3(b) at ~49 MHz and ~91 MHz with these spin transitions, respectively. Note that although -1/2 → +1/2 is an allowed spin transition (dashed arrow in Fig. 3(a)), it does not appear in the ODMR spectrum since there is only a difference in PL intensity between $m_s = \pm 1/2$ and $m_s = \pm 3/2$ doublets [32,45]. The small satellite peaks (~54 MHz, ~84 MHz, and ~96 MHz) occur due to the interaction of V2 centers with a next-nearest-neighbor (NNN) [29]Si nuclei [46]. The remaining peaks are attributed to weaker spin transitions ($\Delta m_s = \pm 2$) or to harmonics of the RF source driving the main two transitions. The ODMR spectrum and contrast are similar to what has been observed for room temperature ODMR with non-resonant laser excitation. For resonant laser excitation, plotted in Fig. 3(c), the ODMR spectrum looks quite similar except that the contrast is much higher. There is also more broadening of the resonances at high RF power than observed for non-resonant laser excitation.

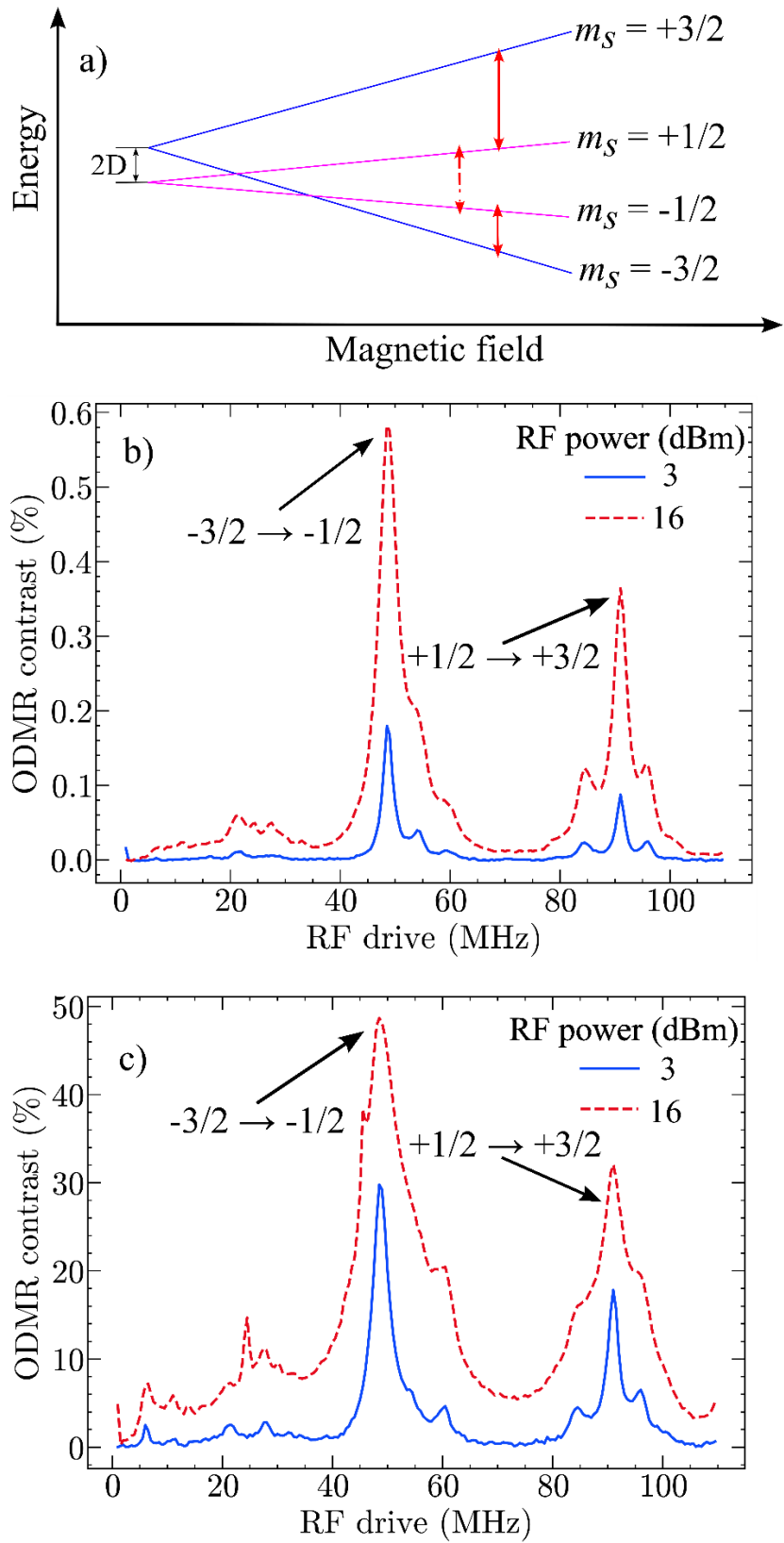

Figure 3: (a) S = 3/2 system of V2 ground state with $\Delta m_s = \pm 1$ transitions. (b) and (c) ODMR plots for off-resonant and resonant optical excitation, respectively for RF powers 3 dBm and 16 dBm. The two most prominent peaks are labeled with corresponding spin transitions.

For further study of ODMR contrast, we focus on the -3/2 → -1/2 transition as it yields the largest ODMR contrast for both resonant and off-resonant excitation. Fig. 4(a) plots the ODMR contrast as a function of laser power. The off-resonant excitation (purple) can reach a maximum ODMR contrast of ~0.6%, which is comparable to room temperature measurement. Moreover, the off-resonant ODMR contrast stays constant even at higher laser powers (through 100 mW). Thus, the off-resonant ODMR contrast does not improve significantly at low temperatures or higher laser powers. For resonant excitation (red), the ODMR contrast reaches a maximum of ~50% for 2 μW of 916 nm laser power. Note that we concurrently supply 2 μW of 785 nm laser during resonant excitation to mitigate photoionization in the system and ensure the correct negative charge $q = -1$ for V2 centers [47]. The power dependences of the PL signals are also plotted in Fig. 4(b). The PL saturates at a few mW for resonant excitation (red), while there is little sign of saturation for non-resonant excitation (purple). The maximum PL signal for resonant excitation is at least two orders of magnitude weaker than for non-resonant excitation. Clearly, the decrease in PL intensity for resonant excitation must be attributed to only a small fraction of the V2 ensemble having its ZPL resonant with the narrow laser line. We hypothesize that the much higher ODMR contrast for resonant excitation is due to spin-selective optical excitation [47], essentially the same as observed for single defects. Considerations for an ensemble will be discussed in Section IV, but this spin selectivity requires the laser linewidth and single defect ZPL linewidths to be less than the 1 GHz splitting between ±1/2 and ±3/2 optical transitions [47].

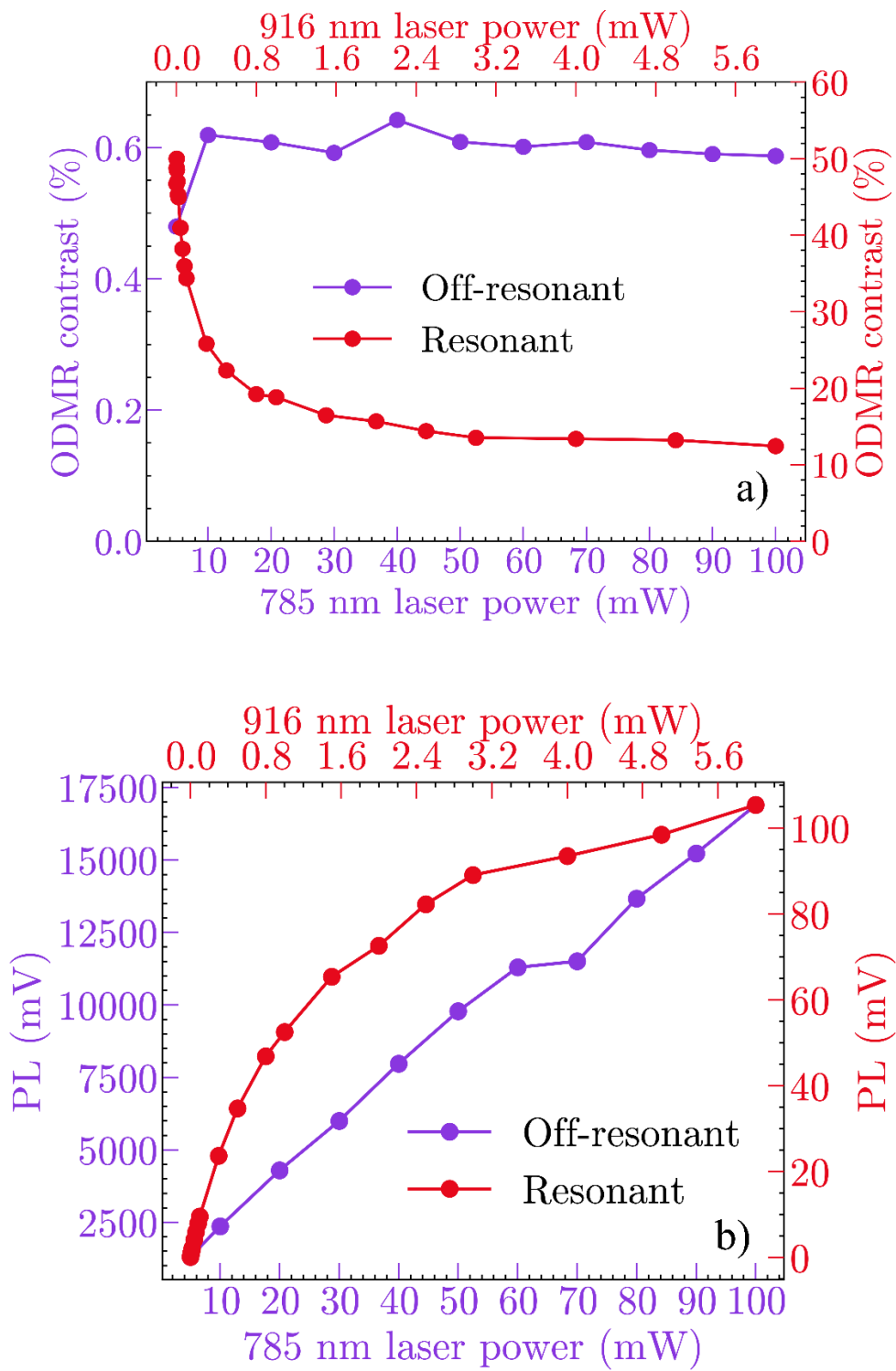

Figure 4: (a) ODMR and (b) PL plots for off-resonant (purple) and resonant (red) laser excitation.

## C.     Effects of temperature and laser modulation on ODMR contrast

To validate our hypothesis for the high contrast, we artificially broaden the tunable laser linewidth to 1 GHz, corresponding to the splitting between the two spin-conserving optical transitions between the ground and excited states [47]. Here, the term "artificial broadening" refers to the rapid modulation of the tunable laser frequency (inset of Fig. 5(a)). Assuming we can scan the laser frequency faster than the relevant polarization and readout timescales, we can mimic the broadening of laser linewidth ($\Delta f$) up to 1 GHz and beyond to drive both of the optical spin transitions and observe if the contrast decreases. Fig. 5(a) plots the ODMR contrast vs the laser modulation amplitude. The laser frequency is sinusoidally varied at a scanning frequency of 5 kHz. As the modulation amplitude increases, the contrast decreases, approaching 10% for $\Delta f \approx 5$ GHz. We might expect a more precipitous drop in contrast at a bandwidth of 1 GHz as both transitions are driven, but there will still be some spin selectivity if one transition is driven more than the other. Nonetheless, the results conform to our hypothesis.

To expand our analysis of resonant ODMR contrast, we vary the sample temperature. At each temperature, the focus is adjusted and the laser is tuned to the maximum PLE. In Fig. 5(b), the contrast decreases from 56% at 4 K to ~0.25% at 60 K, which is similar to the off-resonant contrast. From Ref. [48], we expect that increasing the temperature induces phonon-mediated mixing of polaronic states, leading to the broadening of the optical transition linewidth. As a result, the spin-selective optical transitions start to overlap. Ref. [48] demonstrated a sharp increase in ZPL linewidth at ~25 K for V2 centers in 4H-SiC, with a measured linewidth of ~0.25 GHz at 26 K. This matches up well with our data, where the ODMR contrast sharply decreases at ~20 K.

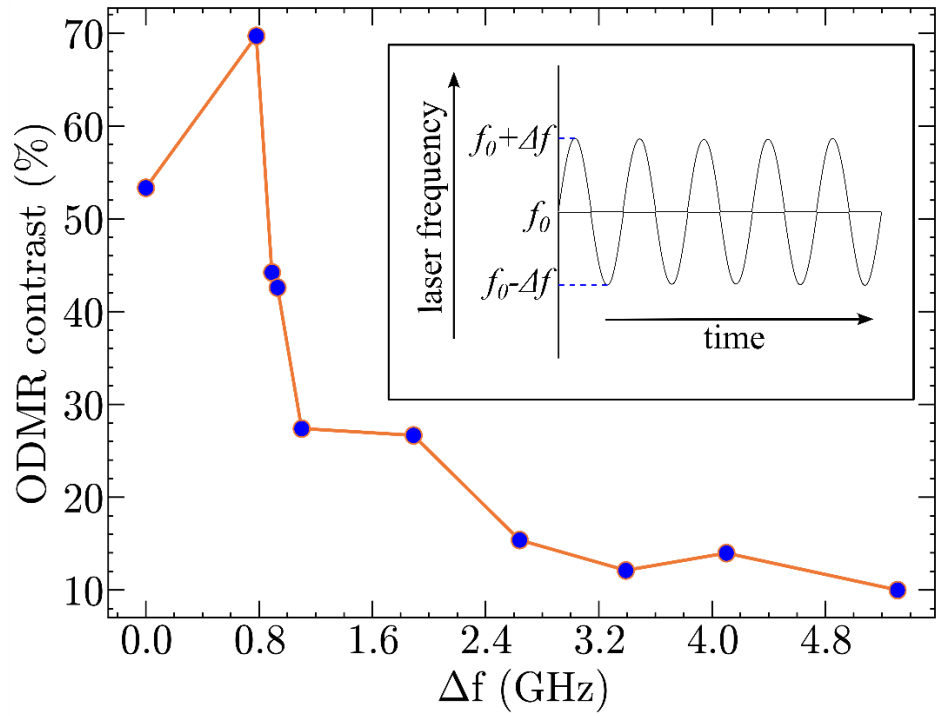

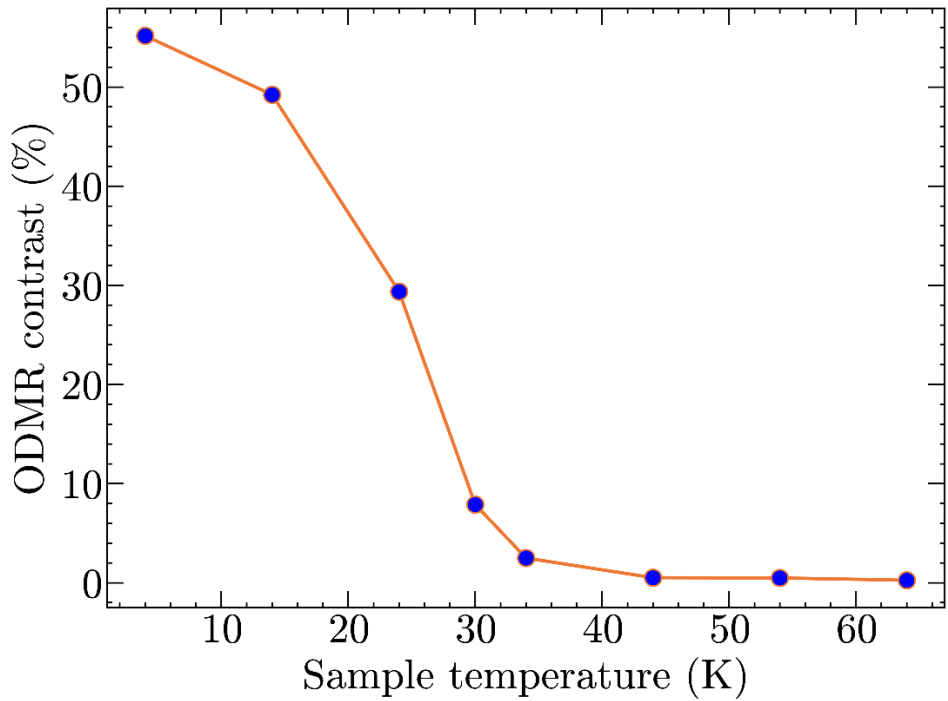

Figure 5: (a) ODMR contrast as a function of laser linewidth modulation. (b) ODMR contrast as a function of sample temperature.

### D.    Calculation of magnetic sensitivity

For the simplest approach to magnetic field sensing, the cw ODMR is measured at the point of maximum slope, $\alpha = dC/df$, where $C$ is the ODMR contrast and $f$ is the RF frequency. We perform ODMR scans for different RF powers and identify the RF power, $RF_{steep}$, that generates the steepest slope for the ODMR peak. For both excitation types, we get maximum slopes at 16 dBm. Then, we set the power to $RF_{steep}$ = 16 dBm and obtain a continuous time-series data with the SR860 lock-in amplifier. The laser powers for resonant and off-resonant excitation were set to 300 μW and 30 mW, respectively. We set the time constant of the lock-in amplifier to 10 ms, and the captured time-series data has a sampling rate of 1.2 kHz with 500,000 sample counts. Afterward, we convert the time-series data into magnetic field fluctuations using the slope and calculate the magnetic field amplitude spectral density through Welch's method. This spectral density is a measure of the magnetic sensitivity. Furthermore, to estimate the shot-noise-limited sensitivity for cw ODMR, we use the following equation [25]:

$$\eta_{cw} = \frac{\hbar}{g_e \mu_B} \frac{1}{\alpha \sqrt{R}} \, , \qquad (2)$$

where, $R$ is the photon detection rate.

Table 1: Summary of magnetic sensitivity (nT/√Hz) for resonant and non-resonant excitation

| Excitation type | Magnetic sensitivity within the low frequency limit (nT/√Hz) | Shot-noise-limited sensitivity (nT/√Hz) |
|---|---|---|
| Resonant | 99.6 | 1.52 |
| Off-resonant | 206 | 1.83 |

Fig. 6 plots the magnetic sensitivity for both excitation types, and Table 1 summarizes the key findings. It is seen that the sensitivity of the resonant excitation is almost twice as low as the off-resonant case within the low frequency limit. Moreover, the acquired sensitivity of the off-resonant system comes at the cost of high laser power: ~100 times more than that of the resonant setup. Therefore, with 300 μW of resonant laser power, we can achieve sub-μT/√Hz sensitivity.

We also calculate the shot-noise-limited sensitivity for resonant and off-resonant excitation, which are 1.52 nT/√Hz and 1.83 nT/√Hz, respectively (plotted as horizontal lines). At higher frequencies, the measured sensitivity gets closer to these values, although some of the decrease is due to the lock-in time constant. The noise peak at ~100 Hz for the off-resonant setup comes from separately measured noise from the 785 nm laser itself.

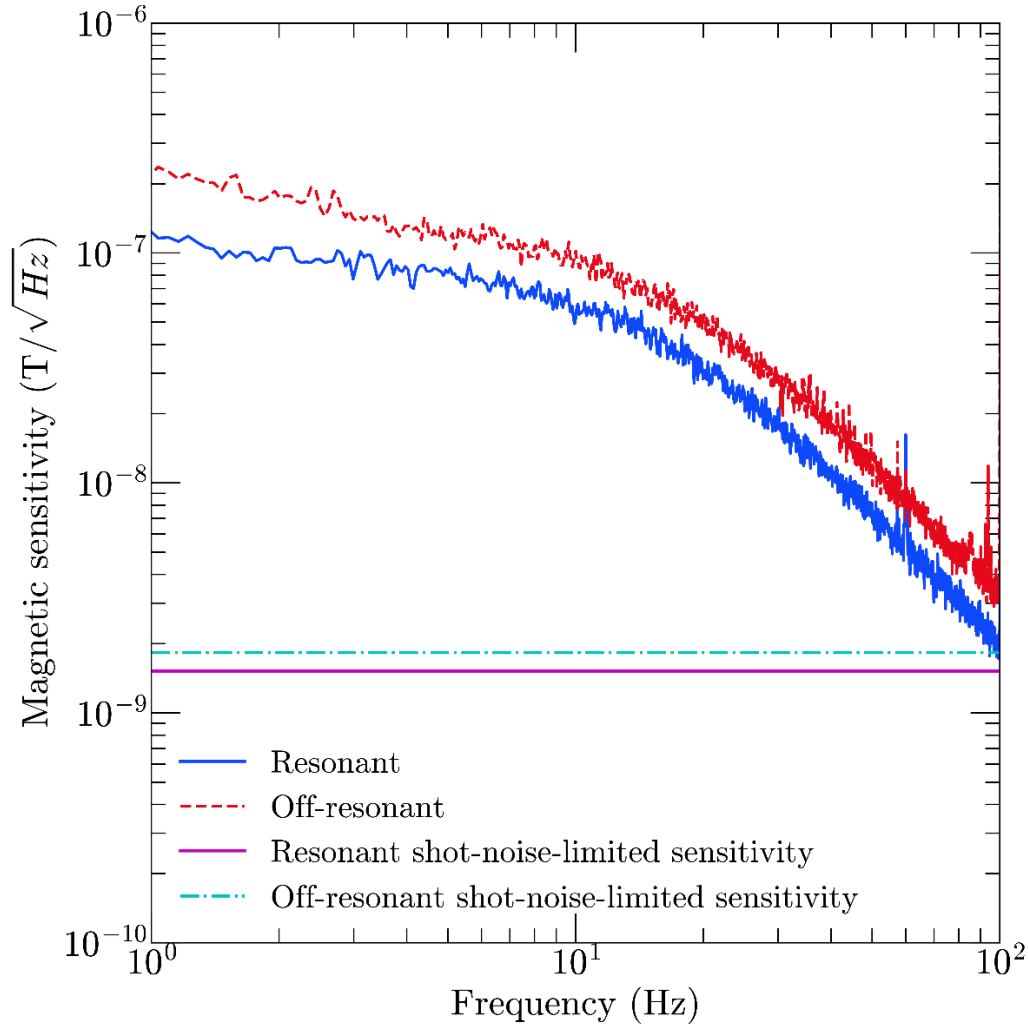

Figure 6: Magnetic sensitivity curves for resonant and off-resonant excitation. The calculated sensitivity for resonant and off-resonant excitation within the low frequency limit are 99.6 nT/√Hz and 206 nT/√Hz, respectively. The horizontal lines represent the shot-noise-limited sensitivity.

## IV. Optical pumping of the sub-ensemble of defects

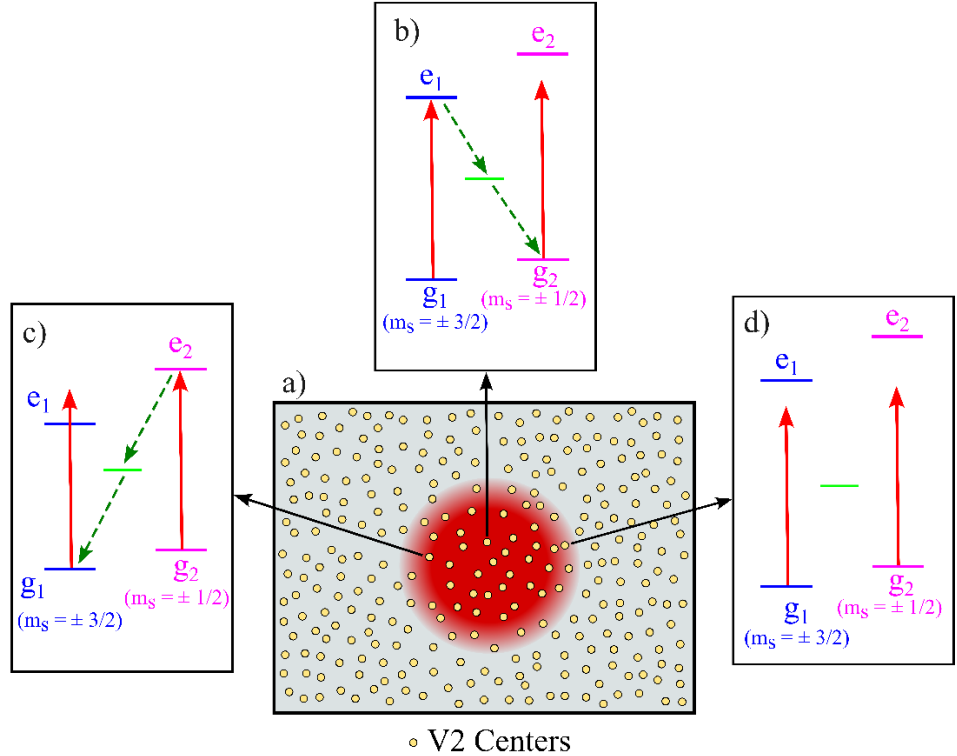

Figure 7: (a) Top view of 4H-SiC sample (gray) with V2 centers (yellow circles). (b) – (d) demonstrate spin-selective optical transitions and optical pumping of different V2 centers within the laser excitation spot (red circle).

To better understand the spin-selective optical transitions and high contrast within the ensemble of V2 centers for resonant excitation, let us consider Fig. 7. As previously mentioned, the narrow linewidth of the resonant laser excites a "subset" of the V2 centers within the ensemble. For a particular resonant laser wavelength, this subset of defects can undergo one of two optical transitions. For instance, Fig. 7(b) and Fig. 7(c) demonstrate the $m_s = \pm 3/2$ states and $m_s = \pm 1/2$ states being resonant with the tunable laser, respectively, for two example V2 centers. When the laser is resonant with $\pm 3/2$ ($\pm 1/2$) transitions, the V2 centers are optically pumped into $\pm 1/2$ ($\pm 3/2$) states through the intersystem crossing (green arrows). For the silicon vacancy, the intersystem crossing is relatively strong for both spin states, which is why non-resonant excitation gives weak contrast. Here, it allows for efficient optical pumping into either set of spin states. After this optical pumping, there is no further optical excitation or PL without spin relaxation or off-resonant excitation of the other transition. This leads to weak PL without RF and a strong increase when driving a transition between ±1/2 and ±3/2. Note this effect does not depend on a net ensemble spin polarization but only on individual V2 centers, resonant with the laser, that are polarized into either spin doublet. As the linewidth of the laser or ZPL linewidth increases, both optical transitions can be driven, leading to lower spin selectivity and ODMR contrast. Within the full ensemble, the ZPLs of most V2 centers will not be resonant at all, as shown in Fig. 7(d). Based on a single defect ZPL linewidth of 0.6 µeV and an ensemble linewidth of 170 µeV (~0.13 nm), roughly 0.6% of V2 centers will contribute to the sub-ensemble. This fraction can be increased by more intense optical driving, but at the expense of spin selectivity due to power broadening.

## V. Conclusion

These results provide an in-depth study of ODMR of ensembles of V2 centers using resonant optical excitation, with an emphasis on potential use in quantum sensing. One of the main results is that at low temperatures of ~4 K, the ODMR contrast can be two orders of magnitude higher with resonant excitation compared to non-resonant, reaching 50%. The dramatic improvement in contrast is attributed to spin-selective optical excitation, in which a sub-ensemble of defects is excited with either the ±1/2 or ±3/2 transitions in resonance with the laser. This selectivity is reduced when the laser linewidth or transition linewidths are increased, such that both transitions are excited. Importantly, this means that the contrast drops with temperature as the ZPL linewidths increase, reaching off-resonant contrast at 60 K. A consequence of only exciting a small sub-ensemble is that the PL is much weaker, counteracting some of the sensing improvement due to better contrast. Still, we demonstrate that through resonant excitation, our low-temperature system can achieve sub-µT/√Hz sensitivity, better than with non-resonant excitation, and requiring orders of magnitude less laser power. Additional gains could be achieved by excitation of a larger portion of the defect ensemble. This could be attained with narrower ensemble ZPL linewidths or using a laser frequency comb that spans more of the ensemble. This resonant excitation setup has the potential to be used in a wide-field magnetic imaging at low temperatures, which is important for characterizing quantum materials and devices. For this imaging, using the edge of a sample is likely impractical, but nanopillars etched into the SiC could provide better coupling of light to the V2 dipole and also improve photon collection and spatial resolution.

## DATA AVAILABILITY

The data that support the findings of this article are not publicly available because the difficulty of preparing, depositing, and hosting the data would be prohibitive. The data are available from the authors upon reasonable request.

## **<u>REFERENCES</u>**